\documentclass[letterpaper, 10 pt, conference]{ieeeconf}  

\IEEEoverridecommandlockouts                              
\usepackage{multirow}
\usepackage[xetex]{graphicx}
\usepackage[export]{adjustbox}
\usepackage{caption}
\usepackage[xcdraw]{xcolor}
\usepackage[%
    colorlinks=true,
    pdfborder={0 0 0},
    linkcolor=blue
]{hyperref}

\usepackage{epsfig} 
\usepackage{mathptmx} 
\usepackage{times} 
\usepackage{amsmath} 
\usepackage{amssymb}  
\usepackage[font=small, skip=1.2pt]{caption}
\title{\LARGE \bf
Deep learning based non-contact physiological monitoring in Neonatal Intensive Care Unit
}
\author{Nicky Nirlipta Sahoo$^{1,*}$,  Balamurali Murugesan$^{1,2}$, Ayantika Das$^{1}$,\\
Srinivasa Karthik$^{1,2}$,  Keerthi Ram$^{2}$, Steffen Leonhardt$^3$, Jayaraj Joseph$^{1,2}$ and Mohanasankar Sivaprakasam$^{1,2}$
\thanks{$^{*}$ sahoonicky@gmail.com}
\thanks{$^{1}$ Indian Institute of Technology Madras (IITM), 600036, India 
        }%
\thanks{$^{2}$ Healthcare Technology Innovation Centre (HTIC), IITM, 600113, India}
\thanks{$^{3}$ Medical Information Technology (MedIT), Helmholtz-Institute for Biomedical Engineering, RWTH Aachen University, Aachen, 52074, Germany
}
}
\begin{document}
\maketitle
\thispagestyle{empty}
\pagestyle{empty}
\begin{abstract}
Preterm babies in the Neonatal Intensive Care Unit (NICU) have to undergo continuous monitoring of their cardiac health. Conventional monitoring approaches are contact-based, making the neonates prone to various nosocomial infections. Video-based monitoring approaches have opened up potential avenues for contactless measurement. This work presents a pipeline for remote estimation of cardiopulmonary signals from videos in NICU setup. We have proposed an end-to-end deep learning (DL) model that integrates a non-learning-based approach to generate surrogate ground truth (SGT) labels for supervision, thus refraining from direct dependency on true ground truth labels.
We have performed an extended
qualitative and quantitative analysis to examine the efficacy of our proposed DL-based pipeline and achieved an overall
average mean absolute error of 4.6 beats per minute (bpm) and root
mean square error of 6.2 bpm in the estimated heart rate.
\end{abstract}

\section{Introduction}
The Neonatal Intensive care units (NICUs) are specialized care units whose vital functionality is to continuously monitor the physiological parameters of neonates. The current gold standard for measuring these parameters is contact-based sensors such as the electrocardiogram (ECG) and photoplethysmogram (PPG). The main hindrance to the utilization of these sensors is the direct skin-contact usage protocol of these devices with adhesive electrodes or transducers. The skin-contact nature of these devices makes the sensitive skin of the neonates infection-prone or even leads to skin damage in case of prolonged usage. Also, these traditional devices need frequent human intervention for calibration. And most importantly, the COVID 19 pandemic has increased the chance of various nosocomial \cite{ramasethu2017prevention} infections. 
Given these concerns, many researchers have started exploring the use of non-contact physiological monitoring equipment. The development of camera-based physiological monitoring, such as remote photoplethysmography (rPPG), presents an opportunity to address the above limitations of the contact-based sensor while providing comfort and safety. The fundamental enabler of rPPG is digital camera sensors which capture the subtle changes in optical properties of skin because of pulsating variations in blood volume.

The task of PPG extraction from images is conventionally solved by traditional multistage pipelines like color space transformation followed by signal decomposition and filtering \cite{villarroel2019non}, \cite{khanam2021non}. These methods have a heavy dependency on pre-processing tasks like region of interest (ROI) detection, skin segmentation, etc. The cumbersome nature of these pre-processing steps has brought in the necessity of automated methods. With the great success of DL for automating various computer vision tasks, DL methods have also been proposed for pre-processing. Villarroel \emph{et~al.} \cite{villarroel2019non}, Chahl \emph{et~al.} \cite{khanam2021non} have adopted DL based pre-processors feeding into traditional pipelines for efficient extraction of vital signals like PPG in NICU setup. However, conventional multistage pipelines for PPG extraction are susceptible to variations in parameters like ROI dimensions considered for the operation, affecting the robustness of the results. This calls for an end-to-end learning framework which is more desirable over these methods.

In order to elevate the generalizing capability of PPG extraction approaches, various DL methods have been proposed like PhysNet \cite{yu2019remote}, Deepphys \cite{chen2018deepphys}. Although such contributions have addressed efficient PPG extraction methods, all of them have experimented with adult subject data-sets acquired in a highly constrained environment. The real-world NICU data presents a lot of challenges, unlike adults, neonates have underdeveloped features and they show an abrupt change in motion. Also, most of these approaches do not account for realistic use cases like remote functionality, where the unavailability of true ground truth PPG waveforms is a more valid assumption. This gap between the methods and their remote real-time applicability has motivated our work herein.

To mitigate the above challenges, we have adopted anatomy agnostic pre-processing steps and SGT extraction methods. We have imparted the anatomy agnostic property by handcrafting suitable subject-specific ROIs from variable anatomies of the neonates in NICU data. The SGT extraction method involved conventional pipelines like CHROM \cite{de2013robust} and POS \cite{wang2016algorithmic}. Although it might apparently seem that the most viable approach for pre-processing is a pre-trained DL-based pre-processing pipeline, the distribution shift between standard rPPG datasets available in the public domain and the NICU-based video data \cite{paul2020non} we have used is quite considerable. 

In this work, we have proposed a method for complete end-to-end PPG waveform extraction. Initiating with only real-time NICU data \cite{paul2020non} in hand and assuming the unavailability of true ground truth, we have brought together suitable SGT extraction methods from the data. Thus establishing a ground for supervision, we have trained PhysNet \cite{yu2019remote}, a DL-based PPG extractor, for robust and efficient performance. 
Our contributions can be summarized as follows:
\begin{enumerate}
    \item An end-to-end DL framework: PhysNet, for PPG extraction in NICU setup, supervised by surrogate ground truth labels extracted using CHROM \cite{de2013robust};
    \item Qualitative and quantitative experiments to check the efficacy of the PhysNet in NICU setup in variable anatomies of the neonates;
    \item Several metrics-based analysis to show superiority in performance of PhysNet as compared to conventional approaches;
    \item Evaluation of PhysNet with a  different camera setup; near infrared (NIR), which was in sync with the primary RGB camera during acquisition.
\end{enumerate}
\section{MATERIALS AND METHODS}
In the following, data set and the algorithms used for direct PPG extraction from the video have been discussed.
\subsection{Data Collection}
\label{subsec}
A single-center pilot study was conducted in April 2018 at Saveetha Medical College Hospital and approved by the institutional Ethics Committee of Saveetha University (SMC/IEC/2018/03/067) \cite{paul2020non}. This study consists of 19 video recordings of different neonatal subjects, each of 10 min duration. The camera set up for the recordings consisted of three sensitive CMOS cameras; one color camera (GS3-U3-23S6C-C; FLIR, USA) and two monochrome cameras (Grasshopper 3 GS3-U3-23S6M-C). All the videos were recorded in Bayer format at 25 frames per second (FPS) with a frame resolution of 1920×1200 and stored in uncompressed format. The standard monitoring device, Radical-7 (Masimo, USA) pulse oximeter is used for reference, which has measurements of pulse rate (PR), perfusion index (PI), peripheral oxygen saturation and PPG signal.
\subsection{Architecture}
\begin{figure}[t!]
\begin{center}
\vspace{2.5pt}
\raggedleft
  
  \vspace{2pt}
  \includegraphics[width=0.485\textwidth,height=0.35\textwidth]{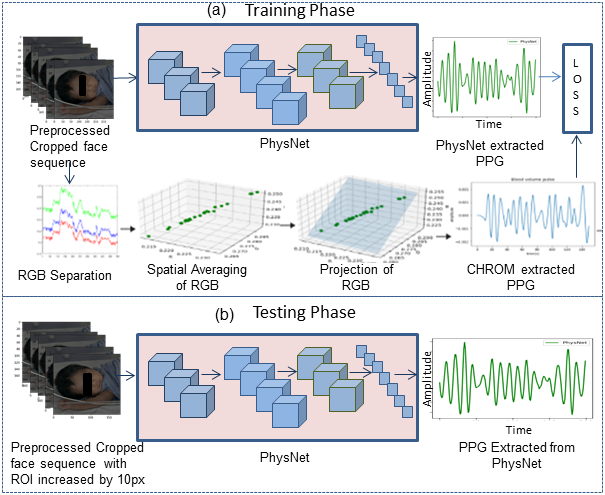}
  \caption{\textbf{Architecture of the proposed network: }\string(a) Training phase and (b) Testing Phase.}
  \vspace{-2.3em}
  \label{architecture}
  \end{center}
\end{figure}

The DL network, PhysNet \cite{yu2019remote} extracts meaningful features directly from videos to learn to regress valid points for PPG waveform estimation, aided by the SGT generators. The network architecture is schematically shown in Fig \ref{architecture}. For SGT extraction CHROM is used, which consists of spatial averaging of RGB signals, temporal normalization for the elimination of DC value and projection of these signals onto a plane such that pulsating signal can be extracted. We leverage the SGT PPG signals extracted for training the DL model. PhysNet is an end-to-end 3DCNN based Spatio-temporal network that can learn temporal contextual information. To enhance the capabilities of  PhysNet, its output is filtered with amplitude-based filtering \cite{villarroel2019non} to form PhysNet-filt. Also, another successor of PhysNet was trained with skin segmented inputs to the model to refine its performance. This refined network PhysNet-sk uses pre-trained weights from \cite{topiwala2019adaptation} for skin segmentation.
\begin{table*}
\vspace{2.45pt}
\caption{Pulse measurement of the intra class subjects}
\label{intraclass}
\resizebox{\textwidth}{!}{%
\begin{tabular}{|c|ccccc|ccccc|ccccc|}
\hline
\multirow{2}{*}{\textbf{Method}} & \multicolumn{5}{c|}{\textbf{MAE}} & \multicolumn{5}{c|}{\textbf{RMSE}} & \multicolumn{5}{c|}{\textbf{Pearson(r)}} \\ \cline{2-16} 
 & F\_RM & F\_MM & Ab & B & SF\_RM & F\_RM & F\_MM & Ab & B & SF\_RM & F\_RM & F\_MM & Ab & B & SF\_RM \\ \hline
CHROM & 4.97 & 4.8 & 0.84 & 12.08 & 7.03 & 5.73 & 6.05 & 0.88 & 13.02 & 8.62 & 0.4 & 0.73 & 0.81 & 0.48 & 0.42 \\ \cline{1-1}
POS & 7.24 & 5.61 & 1.18 & 10.51 & 5.53 & 8.53 & 6.73 & 1.2 & 12.23 & 6.31 & 0.5 & 0.51 & 0.61 & 0.81 & 0.35 \\ \cline{1-1}
PhysNet & 3.34 & 2.44 & 0.89 & 9.51 & 6.37 & 5.08 & 2.56 & 1.02 & 12.66 & 8.53 & 0.64 & 0.83 & 0.83 & 0.55 & 0.54 \\ \cline{1-1}
PhysNet-Filt & 1.66 & 2.44 & 0.56 & 5.16 & 4.07 & 1.83 & 2.56 & 0.66 & 5.36 & 4.87 & 0.91 & 0.83 & 0.91 & 0.79 & 0.5 \\ \cline{1-1}
PhysNet-sk & 2.51 & 1.35 & 0.44 & 3.45 & 9.34 & 3.94 & 1.84 & 0.52 & 3.89 & 13.6 & 0.62 & 0.95 & 0.97 & 0.76 & 0.6 \\ \hline
\end{tabular}
}
\vspace{-1.5em}
\end{table*}

\begin{table}
\caption{Pulse measurement of the inter class subjects}
\label{interclass}
\begin{adjustbox}{max height = 2\textwidth, width=0.48\textwidth}
\begin{tabular}{|c|ccc|ccc|ccc|}
\hline
\multirow{4}{*}{\textbf{Method}} & \multicolumn{3}{c|}{\multirow{2}{*}{\textbf{MAE}}} & \multicolumn{3}{c|}{\multirow{2}{*}{\textbf{RMSE}}} & \multicolumn{3}{c|}{\multirow{2}{*}{\textbf{Pearson(r)}}} \\
 & \multicolumn{3}{c|}{} & \multicolumn{3}{c|}{} & \multicolumn{3}{c|}{} \\ \cline{2-10} 
 & \multirow{2}{*}{F} & \multirow{2}{*}{Ab} & \multirow{2}{*}{SF\_MM} & \multirow{2}{*}{F} & \multirow{2}{*}{Ab} & \multirow{2}{*}{SF\_MM} & \multirow{2}{*}{F} & \multirow{2}{*}{Ab} & \multirow{2}{*}{SF\_MM} \\
 &  &  &  &  &  &  &  &  &  \\ \hline
CHROM & 5.92 & 2.2 & 5.31 & 9.38 & 2.6 & 5.73 & 0.4 & 0.67 & 0.94 \\ \cline{1-1}
POS & 5.93 & 2.65 & 10.33 & 9.84 & 3.01 & 12.54 & 0.5 & 0.8 & 0.83 \\ \cline{1-1}
PhysNet & 6.27 & 0.85 & 7.04 & 9.15 & 1.07 & 9.5 & 0.55 & 0.8 & 0.73 \\ \cline{1-1}
PhysNet-Filt & 3.6 & 0.85 & 4.32 & 4.8 & 1.07 & 4.72 & 0.65 & 0.8 & 0.97 \\ \cline{1-1}
PhysNet-sk & 3.24 & 0.85 & 6.5 & 4.3 & 1.07 & 6.7 & 0.55 & 0.8 & 0.84 \\ \hline
\end{tabular}
\end{adjustbox}
\vspace{-1.5em}
\end{table}
\section{EXPERIMENTS AND RESULTS}
\subsection{Preprocessing}
The raw Bayer videos were demosaiced to form RGB videos to better adapt to the usage conventions and were down-sampled by a factor of three to reduce computational load. The RGB videos were white balanced with gray world assumption \cite{ebner2007color} to standardize the ambient light. The data acquisition environment was fixed, which aided in the selection of specific ROI within a video. The SGT PPG signals were band-pass filtered with a Butterworth filter in a frequency range of 78 to 240 bpm before passing them to the network.
\subsection{Implementation Details}
We have trained the PhysNet by initializing random weights and supervising with Negative Pearson loss \cite{yu2019remote} calculated between the predicted regression output and SGT from CHROM. The training intricacies include 80 maximum epochs, Adam optimizer, a learning rate of 1e-3 and a batch size of 2, with each mini-batch comprising of a sequence of 148 video frames with corresponding label data points. The input to the physnet is ROI images extracted and resized to $128 \times 128$ which is normalized over each mini-batch. The stopping criteria for training were based on low validation loss values and reasonably pulsating PPG output for validation data. The model was developed using PyTorch and the training process was carried out in a workstation using an i9 18 core 192GB CPU and NVIDIA 24GB GPU.
We have selected 17 out of 19 different neonatal subjects available based on perceptual quality. A subject-specific split was considered, where 13 subjects were shown during training (intra class subjects) and the remaining 4 subjects were unseen (inter class subjects). Further, a video-specific split was done, where 60\% frames from each of the intra class subjects were chosen randomly for training and the remaining 40\% frames, along with all the frames of the inter class subjects, were used for validation.

\subsection{Evaluation Metrics}
The predicted PPG signals from PhysNet are post-processed using a first-order Butterworth filter with cutoff frequencies of 1.5-4 Hz. The heart rate (HR) of the predicted signals is calculated with peak detection in the frequency domain using  FFT on Hanning windowed signals \cite{de2013robust} with a window size of 10s . For evaluation, we have computed three standard metrics: Mean absolute error (\emph{MAE}), root mean square error (\emph{RMSE}) and Pearson correlation (\emph{r}) between the calculated HR and ground truth HR averaged over 10s window, from the pulse oximeter as referred in section \ref{subsec}.
\subsection{Results and Discussion}
\subsubsection{Quantitative Analysis}
We evaluated the performance of PhysNet with metrics \emph{MAE}, \emph{RMSE} and \emph{r} by comparing with conventional methods CHROM \cite{de2013robust} and POS \cite{wang2016algorithmic}, as well as post-processed and pre-processed successor of PhysNet: PhysNet-filt and PhysNet-sk. These performance evaluations are tabulated in Table \ref{intraclass} \& \ref{interclass} for both intra and inter class subjects respectively. To comment on the generalizability of the methods, all the inferences were done on ROIs that have 10 pixels height and width increment compared to the ROIs considered for the training protocol.

The quantitative analyses in the table are grouped according to the type of metrics and each group enlists certain anatomical parts for both intra \& inter class subjects. The anatomies considered for intra \& inter class subjects are as follows: face with random movement (F\_RM), face with more melanin (F\_MM), abdomen (Ab), back (B), side face with random movement (SF\_RM), side face with more melanin (SF\_MM). These certain anatomical parts were chosen for representation to highlight the variability of the error differences.

From Table \ref{intraclass} \& \ref{interclass}, it is evident that conventional approaches are quite sensitive and cannot handle variations, while the DL model is able to learn the complex relationship between pixels and underlying physiological changes. Additionally, the judicious choice of the stopping criteria has aided the model not to over-learn the nature of SGT but rather attain better generalization capability. Also, from Table \ref{interclass}, it is clearly inferable that the model can give a comparable performance in inter-class subjects too. PhysNet-filt and PhysNet-sk have outperformed PhysNet due to their enhanced post and pre-processing steps, respectively. 

Additionally, from the variational anatomical study, it is clear that the model is giving minimal error where the face and abdomen part is visible in the ROI. However, the performance has dropped where the ROI has lesser skin area with random motion as in the subject with side face ROI. In PhysNet-sk, subjects with random motion have shown higher error since the skin segmentation was poor for them. This could be attributed to the weights of the pre-trained skin segmentation model trained on relatively stable subjects and not robust to random motion, as seen in the case of neonates.

\begin{figure}[t!]
\begin{center}


  \includegraphics[width=0.985\linewidth]{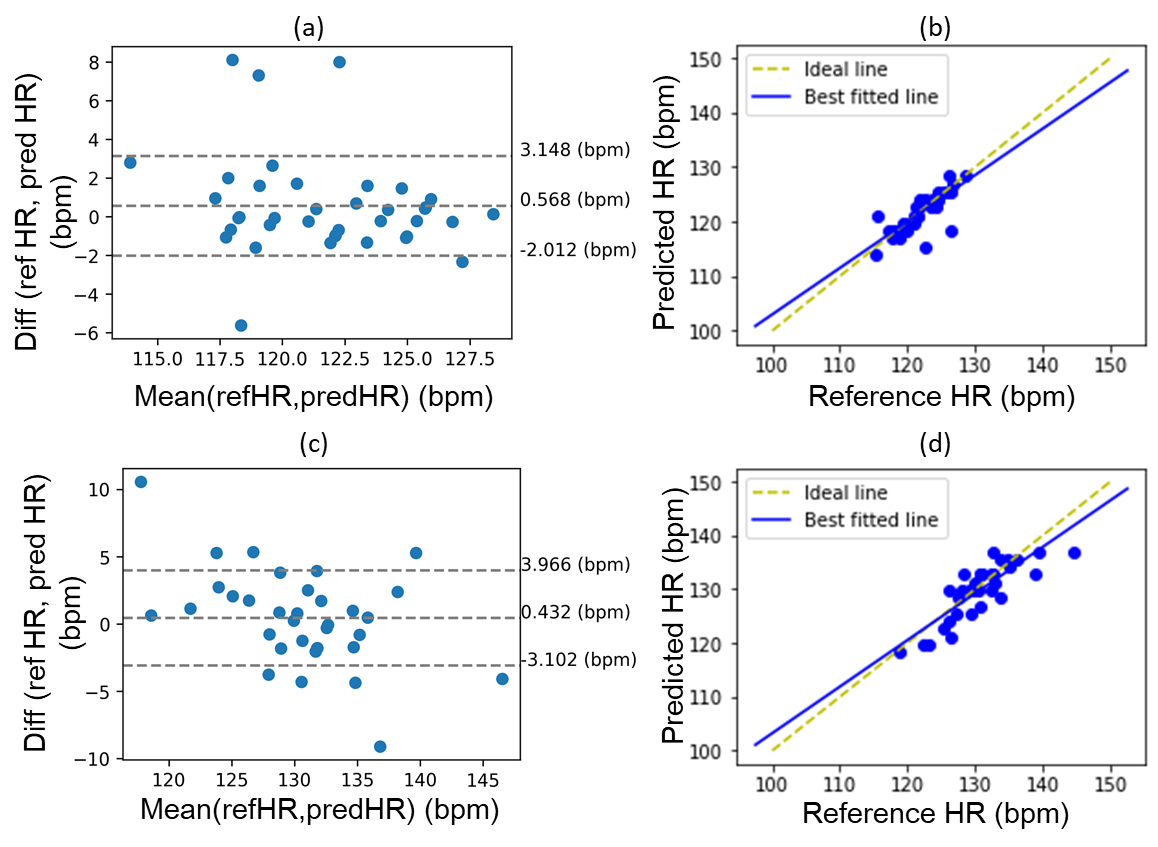}
  
  \caption{\textbf{Comparison between estimated and reference camera heart rate:} (a) \& (b) BA plot and correlation plot for Abdomen (Within training set), (c) \& (d) BA plot and correlation plot for Abdomen (cross validation dataset).}
  \vspace{-2.5em}
  \label{bacorrplot}
  \end{center}
\end{figure}

In order to have a better understanding of the relation between the estimated HR and ground-truth HR, we have shown statistical plots such as Bland-Altman (BA) plots and correlation plots for both intra and inter class in Fig. \ref{bacorrplot}. The BA plot shows an overall mean of around 0.8 bpm and standard deviation of around +4 to -3 bpm which is comparable with results of rPPG in NICU domain \cite{paul2020non}, 
\cite{villarroel2019non}, \cite{khanam2021non}. The correlation plot in Fig. \ref{bacorrplot} shows a strong positive correlation between predicted and ground truth HR with a minimal bias.

\subsubsection{Qualitative Analysis}
\begin{figure}[t!]
\begin{center}
  \vspace{2pt}
  \frame{\includegraphics[width=\linewidth]{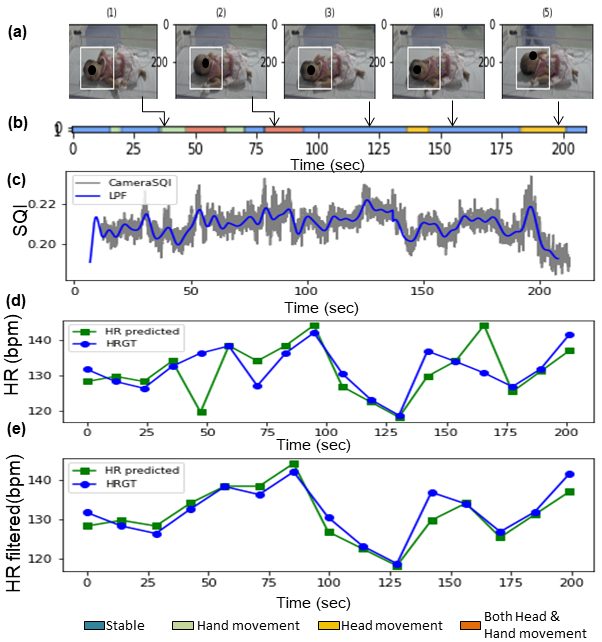}}
  \caption{\textbf{Qualitative Analysis with timing diagram:} (a) Video frames corresponding to time in the timing diagram shown by arrows, (b) Manually annotated timing diagram with color bar representing different activities, (c) Camera SQI for the frames being considered and its low pass filtered signal overlayed, (d) \& (e) Comparison of reference HR and camera derived HR computed using PhysNet and its filtered version respectively.}
  
  \vspace{-1.8em}
  \label{SQI}
    
  \end{center}
\end{figure}

We have analyzed the qualitative performance of the model in a 220s clip of the intra class subject with face ROI (random movements). Fig. \ref{SQI} (a) shows 5 frames with different activities collected at an interval of 40s each \& Fig. \ref{SQI} (c) shows signal quality index (SQI) measured using BRISQUE method \cite{Brisque} capturing the variability in the naturalness of the frame due to motion. Fig. \ref{SQI} (d) \& (e) represents the estimated HR using PhysNet overlaid on the actual HR and its filtered version, respectively. The estimated and actual HR coincide well for most of the video frames, except at timestamps where the subject shows abrupt movements. These outliers are filtered out using the amplitude threshold-based filtering technique \cite{villarroel2019non}, resulting in a correlated HR prediction curve with respect to ground truth HR, Fig. \ref{SQI} (e).

\begin{figure}[t!]
\begin{center}


  \includegraphics[width=\linewidth]{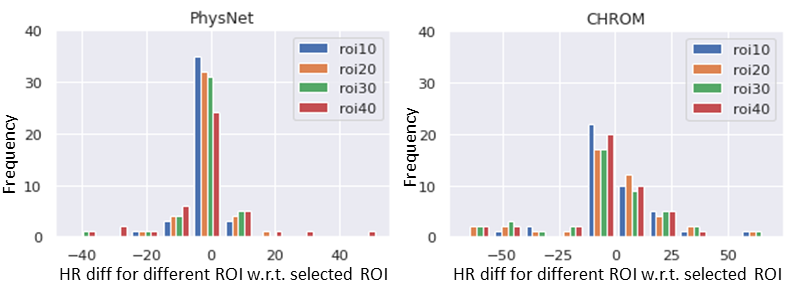}
  \caption{\textbf{Histogram plot} of HR differences between the initial ROI \& the incremented ROIs with their respective frequencies for PhysNet (left) \& CHROM (right) }
  \vspace{-2.5em}
  \label{roihist}
  
  \end{center}
\end{figure} 

\subsubsection{Comparison with different ROI}

To compare the robustness of PhysNet and CHROM with different ROIs, we've examined the relative change in HR of the initial ROI with the HR of dimensionally incremented ROI by 10, 20, 30, 40 pixels in both horizontal and vertical directions. From Fig. \ref{roihist}, it is inferable that PhysNet has a much higher frequency near zero relative change in HR as compared to CHROM, which has a significant deviation from the zero relative change in HR. It's also evident that PhysNet has a lesser spread approximately fitting within a range of $\pm$40 relative change in HR, unlike CHROM, which has a higher spread expanding to a range of $\pm$60 relative change in HR.

\subsubsection{Evaluation on NIR data}
For additional use cases, we extended our experiments performed with RGB cameras to NIR  camera videos using the same training protocol. The NIR and RGB camera being synced in the acquisition setup, the same CHROM extracted PPG signals are used as SGT. Table \ref{NIR} shows comparable performance to that of RGB for both intra class and inter class subjects in the NIR setup.

\begin{table}
\vspace{0.5em}
\caption{Metrics for NIR camera setup}
\label{NIR}
\resizebox{0.48\textwidth}{!}{%
\begin{tabular}{|c|ccc|ccc|ccc|}
\hline
\multirow{6}{*}{\textbf{Method}} & \multicolumn{3}{c|}{\multirow{2}{*}{\textbf{MAE}}} & \multicolumn{3}{c|}{\multirow{2}{*}{\textbf{RMSE}}} & \multicolumn{3}{c|}{\multirow{2}{*}{\textbf{r}}} \\
 & \multicolumn{3}{c|}{} & \multicolumn{3}{c|}{} & \multicolumn{3}{c|}{} \\ \cline{2-10} 
 & \multicolumn{2}{c|}{\multirow{2}{*}{INTRA}} & \multirow{2}{*}{INTER} & \multicolumn{2}{c|}{\multirow{2}{*}{INTRA}} & \multirow{2}{*}{INTER} & \multicolumn{2}{c|}{\multirow{2}{*}{INTRA}} & \multirow{2}{*}{INTER} \\
 & \multicolumn{2}{c|}{} &  & \multicolumn{2}{c|}{} &  & \multicolumn{2}{c|}{} &  \\ \cline{2-10} 
 & \multirow{2}{*}{F} & \multicolumn{1}{c|}{\multirow{2}{*}{F\_MM}} & \multirow{2}{*}{SF} & \multirow{2}{*}{F} & \multicolumn{1}{c|}{\multirow{2}{*}{F\_MM}} & \multirow{2}{*}{SF} & \multirow{2}{*}{F} & \multicolumn{1}{c|}{\multirow{2}{*}{F\_MM}} & \multirow{2}{*}{SF} \\
 &  & \multicolumn{1}{c|}{} &  &  & \multicolumn{1}{c|}{} &  &  & \multicolumn{1}{c|}{} &  \\ \hline
\multirow{2}{*}{PhysNet} & \multirow{2}{*}{6.39} & \multicolumn{1}{c|}{\multirow{2}{*}{3.75}} & \multirow{2}{*}{6.41} & \multirow{2}{*}{8.7} & \multicolumn{1}{c|}{\multirow{2}{*}{5.29}} & \multirow{2}{*}{8.65} & \multirow{2}{*}{0.5} & \multicolumn{1}{c|}{\multirow{2}{*}{0.77}} & \multirow{2}{*}{0.77} \\
 &  & \multicolumn{1}{c|}{} &  &  & \multicolumn{1}{c|}{} &  &  & \multicolumn{1}{c|}{} &  \\ \cline{1-1}
\multirow{2}{*}{PhysNet-Filt} & \multirow{2}{*}{4.96} & \multicolumn{1}{c|}{\multirow{2}{*}{2.67}} & \multirow{2}{*}{3.85} & \multirow{2}{*}{6.74} & \multicolumn{1}{c|}{\multirow{2}{*}{3.52}} & \multirow{2}{*}{5.34} & \multirow{2}{*}{0.64} & \multicolumn{1}{c|}{\multirow{2}{*}{0.9}} & \multirow{2}{*}{0.98} \\
 &  & \multicolumn{1}{c|}{} &  &  & \multicolumn{1}{c|}{} &  &  & \multicolumn{1}{c|}{} &  \\ \hline
\end{tabular}%
}
\vspace{-2em}
\end{table}

\section{CONCLUSION}

We have demonstrated the performance of PhysNet for remote video-based PPG extraction in NICU setup. Our experiments have shown that PhysNet has outperformed conventional approaches for flexible ROI inputs, which is attributed due to the robustness of DL methods and the precise choice of the stopping criteria. We have also evaluated the effectiveness of PhysNet in variable anatomical regions and a different camera setup (NIR). Through this experimental process, we have opened avenues for video-based PPG extraction using DL-based approaches in real-time setup using SGT. Future work can be propagated by deploying more efficient methods to make the DL model self-supervised by eliminating the dependency on conventional methods and the series of post and pre-processing steps involved. This can impart better generalizability and reduce latency.

\addtolength{\textheight}{-12cm}   




\section*{ACKNOWLEDGMENT}
The authors would like to thank the clinical staff at Saveetha Medical College Hospital for their cooperation and assistance in collecting the neonatal videos.




{\small
\bibliographystyle{plain}
\bibliography{root}
}
\end{document}